# Tunable terahertz reflection of graphene via ionic liquid gating

**Yang Wu[1,2,3], Xuepeng Qiu[1], Hongwei Liu[3], Jingbo Liu[4], Yuanfu Chen[4], Lin Ke[3], and Hyunsoo Yang[1,2]**

*1. Department of Electrical and Computer Engineering, National University of Singapore, 117576, Singapore*
*2. Centre for Advanced 2D Materials, National University of Singapore, 6 Science Drive 2, 117546 Singapore*
*3. Institute of Materials Research and Engineering (IMRE), 3, Research Link, 117602, Singapore*
*4. State Key Laboratory of Electronic Thin Films and Integrated Devices, University of Electronic Science and Technology of China, Chengdu 610054, China*

E-mail: eleyang@nus.edu.sg

## Abstract

We report a highly efficient tunable THz reflector in graphene. By applying a small gate voltage (up to ± 3 V), the reflectance of graphene is modulated from a minimum of 0.79% to a maximum of 33.4% using graphene/ionic liquid structures at room temperature, and the reflection tuning is uniform within a wide spectral range (0.1 – 1.5 THz). Our observation is explained by the Drude model, which describes the THz wave-induced intraband transition in graphene. This tunable reflectance of graphene may contribute to broadband THz mirrors, deformable THz mirrors, variable THz beam splitters and other optical components.

---

## 1. Introduction

Recently, many unique properties of THz waves have been intensively explored, indicating great promise for a wide range of applications [1, 2]. For example, THz waves offer a good balance between the penetration depth and resolution in imaging [3, 4]. For spectroscopy related applications, it has been found that many chemical bonds show fingerprints in the THz range, making THz technology intrinsically suitable for the detection of certain materials [5, 6]. Furthermore, THz waves also show its potential for high speed communications [7]. For these THz applications, high quality optical components are of great importance. Moreover, it is also highly desired that the properties of these components can be tuned efficiently [8-12].

Enormous attention has been paid to graphene since it was discovered [13]. Due to the linear band structure of graphene [14], the charge carrier concentration is effectively tunable by a gate voltage [15], consequently contributing to a tunable electrical conductivity with a large range [13, 16-18]. In contrast to interband transitions in the visible and NIR range [19], the optical properties of graphene in the THz range are determined by intraband transitions, due to the low photon energy (4 meV for 1 THz). The THz optical properties can be described by the Drude model $\sigma_{op} = \sigma_{DC}/(1-i\omega\tau)$, where $\tau$ is the Drude scattering time, $\sigma_{op}$ is the optical conductivity and $\sigma_{DC}$ is the optical conductivity in the DC limit. As the scattering rate ($1/\tau$ = 3 to 6 THz) [20] of the graphene carriers is larger than the frequency of our study, it is reasonable to approximate the real part of the THz optical conductivity with the DC conductivity in graphene. With the aim of harnessing these unique properties, graphene has been widely studied in THz devices, such as phase shifters [11, 21], absorbers [22], and modulators [9, 10, 23-26]. However, though the reflectance of graphene has been theoretically predicted to be tunable from 0 to 100% [27] and the reflectance of graphene has been recently tested in the mid infrared and THz range [28], experimental reports on the largely tunable THz reflection of graphene are still missing.

In this work, we experimentally demonstrate the giant tunable reflectance of graphene. In the vicinity of graphene's Dirac point, a minimum reflectance of 0.79% is obtained, in contrast to the maximum reflectance of 33.4% at a gate voltage of 3 V. The spectrum shows that the devices are tuned uniformly over a wide frequency range (0.1 - 1.5 THz). The tuned reflectance range is enormous considering the ultra-thin dimensions of a mono-layer graphene film. The sandwich device structure is compact and simple in geometry. Moreover, all the measurements were performed at room temperature. The low operation voltage of our devices arises from the high gating efficiency of the ionic liquid, in which ions accumulate at the surfaces of the graphene films, and induce the change of the Fermi level of graphene. As the built-in electrical potential drops over only a few nanometres across the interface between graphene and the ionic liquid, a very strong electrical field is formed in graphene, resulting in a high gating efficiency.

## 2. Experiments

Chemical vapour deposition (CVD) graphene was used in this study [29, 30]. As CVD graphene can be fabricated over large areas, the THz beam spot could be fully covered by graphene. Each mono-layer graphene was grown on copper foil, and subsequently transferred onto a quartz substrate. The film quality was inspected using confocal Raman spectroscopy, and the results show prominent G and 2D peaks, while the D peak is negligibly small, indicating the high quality of the graphene films [31]. A photoconductive antenna based THz system is used in this study. To improve the signal to noise ratio, each reported curve in this work was averaged over 300 spectra with a resolution of 7.5 GHz. The system offered a bandwidth from 0.1 to 1.5 THz with a decent signal to noise ratio.

The devices were measured using a reflection configuration, as shown in Fig. 1(a). The sandwich structure is quartz (400 μm)/mono-layer graphene/ionic liquid (300 μm) (1-ethyl-3-methylimidazolium bis (tri-fluoromethylsulfonyl) imide ([EMIM][TFSI])) [32, 33]/mono-layer graphene/quartz (400 μm). High quality electrical contacts, consisting of Cr (10 nm)/Au (90 nm), were deposited on the graphene films by thermal evaporation in a vacuum chamber. The graphene sheets are separated by 300 μm thick spacers to define the cell thickness while avoiding the overlap of reflected peaks. Here, graphene plays the key role for THz reflectance control, while also functioning as the gating electrode [11]. When a positive gate voltage was applied to the device, as shown in Fig. 1(a), negative (positive) ions accumulated in the vicinity within a few nanometers of the top (bottom) graphene layer (Fig. 1(b)), thus inducing a large amount of positive (negative) charges in the top (bottom) graphene film. The main reflection peaks are labelled by different colors in Fig. 1(b) and 1(c): blue for R1, green for R2 and orange for R3, corresponding to the air/quartz, quartz/graphene/ionic liquid and ionic liquid/graphene/quartz interfaces, respectively. Other peaks exist in the time domain data, but can be neglected due to their low amplitudes. Typical time domain results in Fig. 1(c) consist of three peaks, as illustrated in Fig. 1(b). Due to the beam path differences, these three peaks are separated in the time domain.

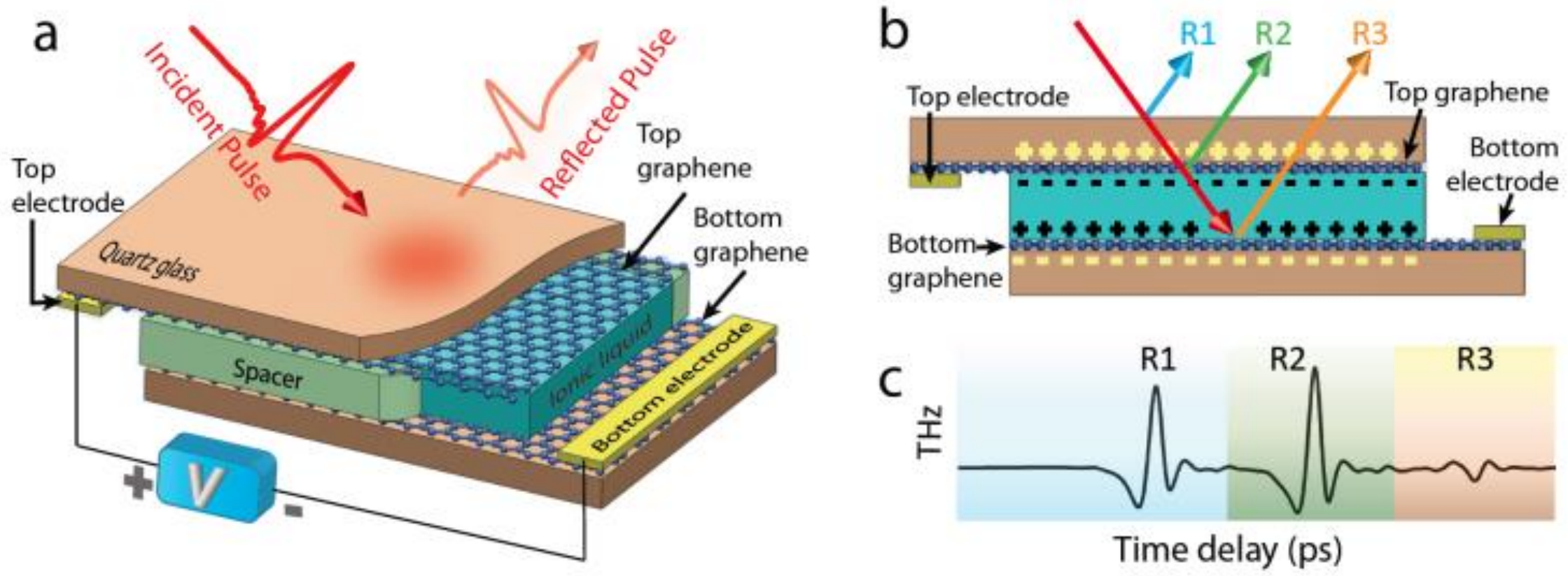

Figure. 1. (a) Ionic liquid cells are sandwiched between two graphene films, which are supported by pieces of quartz. Spacers are used to control the ionic liquid cell thickness, and Cr/Au electrodes are deposited on the graphene films for the application of gate voltages. THz pulses are incident at an angle of ~45° and the reflected pulses are characterized. (b) When a positive voltage is applied to the device, as shown in (a), negative (positive) ions accumulate in the vicinity of the top (bottom) graphene layer, inducing positive (negative) charges in the top (bottom) graphene film. Incident pulses are reflected from the quartz (R1, in blue), top graphene layer (R2, in green) and bottom graphene layer (R3, in orange). (c) A typical THz pulse in the time domain for R1, R2 and R3, while all other peaks are neglected due to their low intensities.

## 3. Results and discussion

Complete THz time domain data are presented in Fig. 2(a), where the x-axis is the time delay in picoseconds, the y-axis is the gate voltage ($V_g$) and the filled colours show the THz E-field amplitude. This contour map shows each time domain curve under various gate voltages. For further analysis, we truncate the time domain results for each reflection peak and display them in Fig. 2(b-d) individually. The time delay (for each peak) induced by the gate voltage is negligible and the curves are shifted along the x-axis for clarity, to show the sequence of applied gate voltages (indicated by the red arrows).

The R1 peak intensity shown in Fig. 2(b) is independent of gate voltage. This is reasonable as the gate voltage does not modify the air/quartz interface. In fact, the uniform pulse shape and amplitude of R1 over the entire range of applied gate voltages attest the high stability of the THz system. The peak amplitudes of R2 shown in Fig. 2(c) increase as the gate voltage increases, which corresponds to hole doping in the top graphene layer. Under the application of a negative gate voltage, the THz reflection first decreases, and then increases. This is because the CVD graphene films are originally p-doped, and the negative voltage pushes the Fermi level of graphene to the charge neutrality point (CNP), which corresponds to both the minimum charge carrier density and the minimum optical conductivity. Next, a further increase of the magnitude of negative gate voltage increases the accumulation of electron charges in graphene, thus increasing the charge carrier concentration as well as the THz reflection.

The peak amplitude of R3 shown in Fig. 2(d) is much smaller than that of R1 and R2 due to the heavy attenuation after many interfaces. Nonetheless, the peak intensity change of R3 is still observable when the gate voltage changes. In comparison to Fig. 2(c), the CNP appears under the application of positive gate voltage, as opposed to that of R2. This interesting behavior can be understood by referring to Fig. 1(b). The top and bottom graphene layers are gated with opposite electric fields in this device structure. As a result, the reflectance tuning shows opposite behaviors for the two graphene layers.

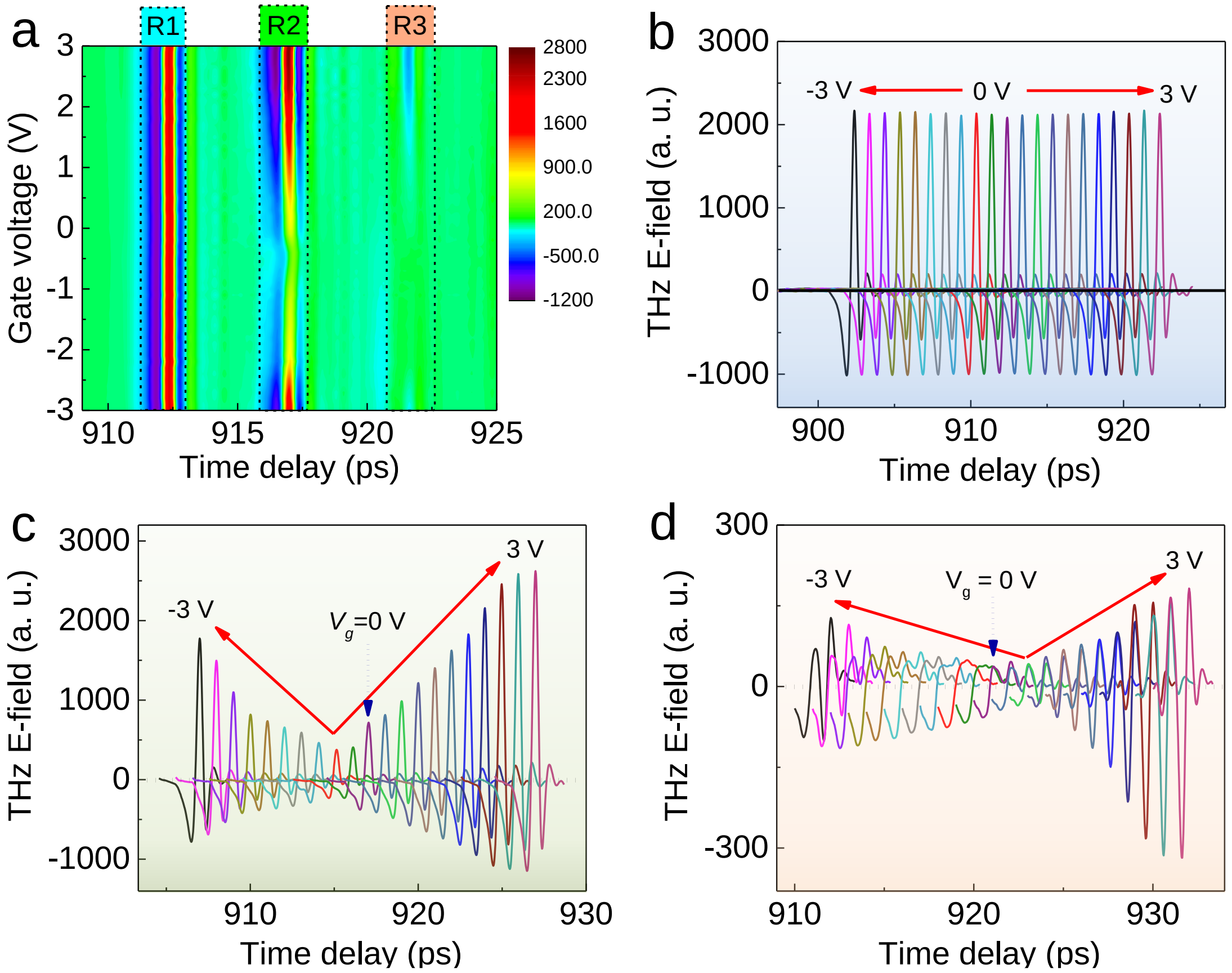

Figure. 2. (a) The contour image of the measured THz data in the time domain, where the x-axis is the time delay, the y-axis is the applied gate voltage (-3 to 3 V), and the filled colours show the E-field strength of the THz pulses. The three peaks are indicated by R1, R2 and R3. (b) The data are truncated from (a) between 908 and 914.4 ps for the analysis of only R1. There are negligible phase shifts introduced by the gate voltage, and the curves are shifted horizontally for clarity. (c) The data are truncated between 914.5 and 918.6 ps for the analysis of only R2, and the $V_g$ = 0 V curve is indicated by the dashed arrow. (d) The data truncated from between 919.9 and 923.2 ps for the analysis of only R3.

Fast Fourier transform (FFT) was performed on the R1 and R2 time domain results. In Fig. 3(a), all the R1 curves overlap between 0.1 to 1.5 THz, and this confirms that the R1 peaks do not change with the gate voltage. The inset of Fig. 3(a) shows the time domain data from the top surface of quartz (red) and a THz gold coated mirror (black) at the same location. The mirror is assumed to be a perfect THz reflector (the reflectance is > 95%). The reflectivity of quartz ($r_1$) is $E_q/E_{m1}$, where $E_{m1}$ is the reflected peak intensity from the mirror at the R1 position, and $E_q$ is that from the quartz.

Reflectance ($R_1$) is $r_1^2$. By this simple calculation, the reflectance of the quartz is about 10% for THz waves, which is in agreement with previous studies [11].

Figure 3(b) shows the FFT results for the R2 time domain data in Fig. 2(c). The reflected THz signal increases for all frequencies within the 0.1 to 1.5 THz range when the gate voltage increases, as indicated by the red dashed arrow. Although our experiment demonstrates the tunable reflectance of graphene only up to 1.5 THz, the device is expected to work for the whole optical range where the photon energy is smaller than $2|E_F|$, where $E_F$ is the energy of the Fermi level. The inset of Fig. 3(b) shows the minimum and maximum THz signals ($E_r$), as well as the signal measured from a mirror at the optical position of R2 ($E_{m2}$). Regardless of the losses in air and quartz, $(1-R_1)\times R_2\times(1-R_1)=(E_r/E_{m2})^2$, where $R_2$ is the reflectance from the top graphene layer interface (R2). From this estimation, $R_2$ is found to be 33.4% at a gate voltage of 3 V, and 0.79% at -0.3 V. This reflectance is substantial considering that it can be attributed to only a single layer of graphene.

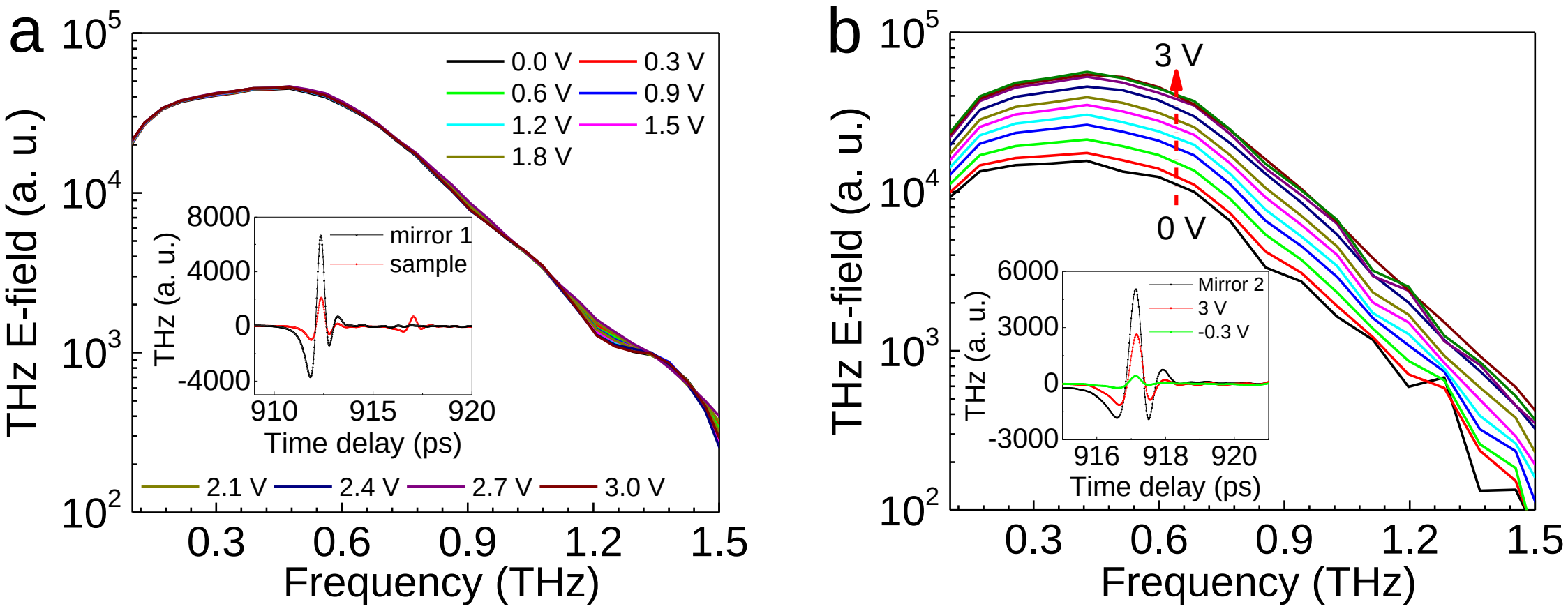

Figure. 3. (a) THz E-field strength from 0.1 to 1.5 THz for R1 at different gate voltages. The inset shows the reference THz signal (black), which was measured from a gold mirror at the same position as the interface corresponding to R1, while the red curve is a curve from Fig. 2(b). (b) The THz E-field strength for R2 at different gate voltages in the frequency domain (0.1 to 1.5 THz). The inset shows three curves (black for mirror, red for $V_g$ = 3 V, and green for $V_g$ = -0.3 V), which were measured at the position of the interface for R2.

As mentioned earlier, the electrical tunability of the graphene reflectance is derived from the tuning of the Fermi level of graphene. In Fig. 4(a), the THz E-field strength of the R2 signal is normalized to the data at $V_g$ = 0 V, and then averaged over 0.1 to 1.5 THz. Four schematic diagrams of the Fermi level of graphene are shown in Fig. 4(a) with labels. For the schematic diagram of No. 1 (No. 4), high negative (positive) voltage is applied, the top graphene layer is highly n-doped (p-doped), therefore, graphene has a high conductivity due to the large number of available density of states. The schematic diagram No. 2 corresponds to the CNP, where the graphene film has the lowest charge carrier density, minimum conductivity, and the lowest THz reflectance in Fig. 2(c). Finally, the schematic diagram No. 3 corresponds to $V_g$ = 0 V, where, due to the intrinsic p-doped character, the Fermi level of graphene is slightly lower than the CNP.

The reflectance of graphene films in the THz range can be explained by the Drude model. Using an optical transfer matrix, the absorption, transmission and reflection of a graphene film can be related to its conductivity [27, 34], and more details can be found in the supplementary data. As shown in Fig. 4(b), the simulated reflection of a graphene film increases monotonically with the increase of its conductivity, which supports our experimental observation, and it can also be further enhanced with an improvement of graphene quality.

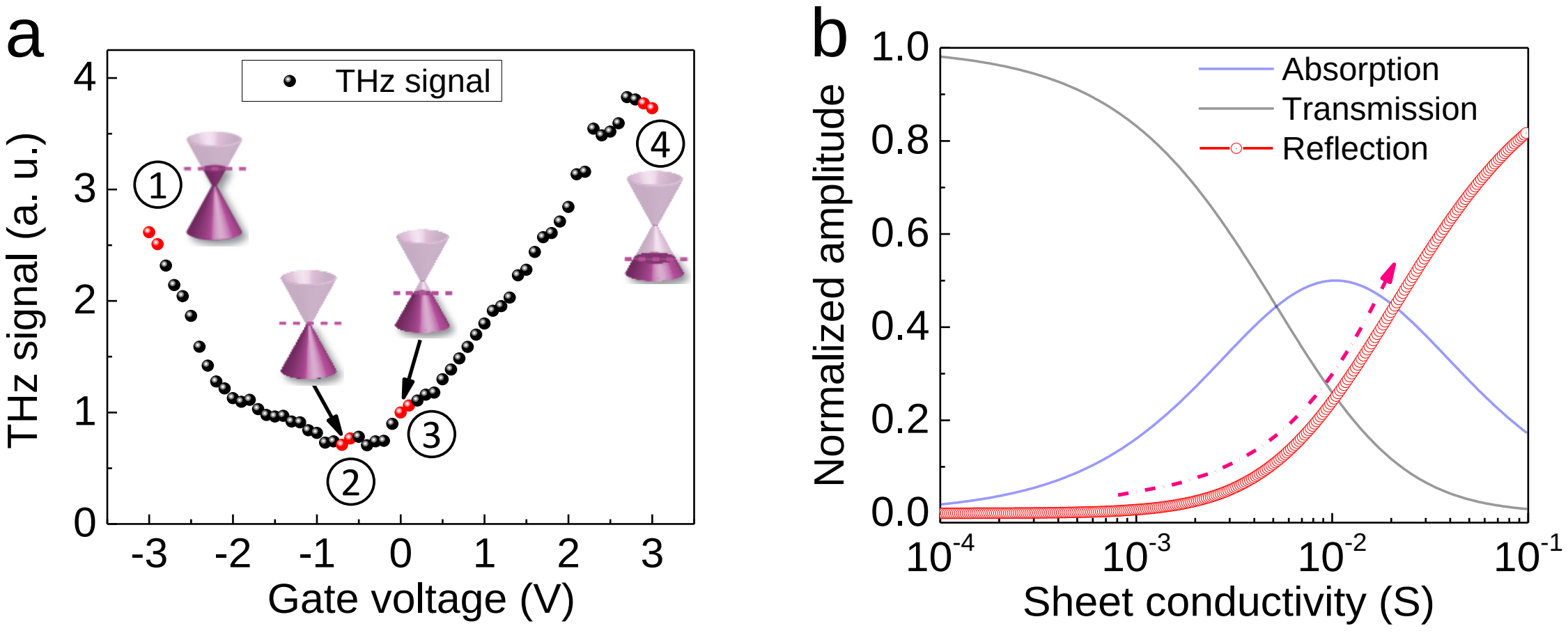

Figure. 4. (a) Normalized THz signal as a function of the gate voltage. Schematic diagrams of the Fermi level in graphene are inserted for four states ($V_g$ = -3, $2V_{Dirac}$, 0, and 3 V), labelled as 1, 2, 3, and 4, respectively. (b) The simulated relationship between reflection/transmission/absorption and graphene sheet conductivity, where the reflectivity is highlighted and its trend is indicated by the red arrow.

In conventional THz mirrors, the thickness of a metal coating layer is required to be at least twice the skin depth *δ*, which is approximately 65 nm for a silver THz mirror [8]. Surprisingly, our study demonstrates that merely a single layer of graphene film can yield substantial reflection, which is promising for THz applications. Furthermore, the ionic liquid-based electrical gating enables spatially resolved THz engineering by simply patterning the graphene films [12]. Analogous to the advent of liquid crystals in electronics [35-37] over the past few decades, it is foreseeable that the devices proposed in this study will similarly pave the way for graphene applications in the THz range. Many devices, such as THz mirrors, deformable THz mirrors/modulators, variable THz beam splitters and even THz 3D imaging will benefit from the tunable reflectance of graphene.

## 4. Conclusion

The ionic liquid gating on graphene films offers a large range tuning of the reflectance of graphene. This is enabled by the unique conical band structure and high carrier mobility of graphene. For a reflectance of 33.4%, the required voltage is only 3 V, which is a consequence of a high gating efficiency of ionic liquid. Furthermore, transfer matrix calculations based on a Drude model can explain the mechanism of this tunability. Our study of the tunable reflectance of graphene could lead to the utilization of other 2D Dirac materials for a wide range of THz applications, and subsequently enable the advanced control of THz waves.

## Acknowledgments

This work is supported by National Research Foundation, Prime Minister's Office, Singapore under its Medium Size Centre programme. The authors thank N. Zhang for Raman spectroscopy characterization.